\documentclass{edi}

\usepackage{makeidx}    
\usepackage{graphicx}   
\usepackage{listings}   
\usepackage{setspace}   
\usepackage{hyperref}   

\makeindex              

\begin{document}
\doublespacing
\chapter{}
\date{}
\title{Diverse End User \\Requirements}

\author{John Grundy, Tanjila Kanij, Jennifer McIntosh, Hourieh Khalajzadeh and Ingo Mueller}
%
%
\maketitle

Software is designed and developed primarily to serve human needs. However, many software systems continue to fail to take into account diverse end users' characteristics, causing frustration, errors and even potentially life threatening situations. These end user human-centric aspects include, but are not limited to, diverse age, ethnicity, gender, personality, cognitive style, language, culture, physical and mental challenges, emotional reactions, socio-economic status, etc.   Software applications need to cover many if not all of these end user human-centric aspects in order to provide a suitable interface, workflow and solution for diverse end users. 

There may be a number of reasons software engineers do not sufficiently take their end user human-centric aspects into account. This includes poor understanding of user needs,  inappropriate designs and time pressures \cite{wirtz2009age,Kavcic2005access,Coiera2011jamia,grundy2020human}. Some larger organisations have dedicated UX/UI and/or customer experience teams that separate developers from end users \cite{da2013understanding}. Many companies are very small and developers need to do all such work themselves, but lack sufficient training in UX, participatory design or other human-centric design methods \cite{ovad2015teaching,nguyen2019closing,shinohara2018teaches}. Software developers are generally well-educated, relatively young, mostly male, most well-conversant in English, of high socio-economic status, and are very comfortable with technology. Because of this, they may find it difficult to empathise, understand and subsequently
incorporate diverse human-centric aspects during the software engineering process~\cite{hartzel2003self,miller2015emotion,stock2008evaluation,grundy2020human}.

As part of our larger research effort to improve support for diverse end user human-centric aspects during software development, we wanted to better understand how developers currently go about addressing these challenging human-centric aspects of their end users in contemporary software development projects.  We wanted to find out which are the key end user human-centric aspects that software developers currently find challenging to address, and how they currently go about trying to address diverse end user human-centric aspects. We wanted to find out what sorts of end user human-centric aspects they tend to encounter, which ones they view as more important and which more challenging to address, what techniques (if any) they currently use to address (some of) them, and where they perceive further research in this area could be done to provide them practical support. 
To this end we carried out a detailed online survey of developers and development team managers, receiving 60 usable responses. We interviewed 12 developers and managers from a range of different practice domains, role specialisations and experience levels to explore further details about issues.

\section{Human Aspects of Users}
\label{sec:background}

Below are some, but by no means all, end user human-centric aspects that software teams need to consider:

\textbf{\textit{Gender:}} Several prominent mainstream articles and books have highlighted gender bias in various technologies, including apps and smart living technologies \cite{perez2019invisible,strengers2020smart}. Recent work has investigated how software, and other systems, are gender biased in various ways \cite{burnett2016gendermag}. 

\textbf{\textit{Age:}} Many smart living systems focus on supporting ageing people. Many educational software systems are targeted to support young children \cite{grundy2018supporting,nouwen2015value}. People of differing ages may have quite different expectations, challenges and reactions to the same software, that need to be addressed \cite{williams2013considerations, agemag}.

\textbf{\textit{Ethnicity and Culture:}} Software that fails to take into account or is biased in terms of ethnicity of people is highly problematic, especially for many emerging smart city applications e.g. policing and surveillance \cite{garvie2016facial}. 

\textbf{\textit{Physical/Mental Challenges:}} Many people live with mental health challenges, cognitive impairment and a wide variety of physical challenges, e.g. impaired mobility, sight, hearing, and speech \cite{stock2008evaluation,zhao2020seenomaly}. Many software solutions have been developed to assist with these challenges, or to take account of them to increase accessibility to software \cite{carcedo2016hapticolor,sierra2012designing}. 

\textbf{\textit{Language:}} Different users speak different languages, have different educational attainment levels, specific colloquialisms and jargon, and different language competencies. Considering these aspects is particularly important during dialogue design, including multi-lingual software and software that adapts to different user dialogue preferences \cite{roturier2015localizing}.

\textbf{\textit{Human Values:}} Values, e.g. inclusiveness, equality, privacy, openness, etc. reflect how, why and to what degree humans value people, objects and ideas \cite{winter2018measuring}. Many apps conflict with one or more human values, causing expectation mismatches and reducing app usage, take-up and acceptance \cite{obie2020first}.

\textbf{\textit{Emotions:}} Different people react differently to technology solutions from an emotional perspective. This includes positive reactions e.g. to a smart home solution providing a feeling of safety, to negative reactions to the same software e.g. feeling lack of control or being monitored intrusively~\cite{curumsing2019emotion}. 


\textbf{\textit{Engagement and Entertainment:}} 
Some people are highly driven by enjoyment, entertainment and `fun' aspects of using software -- computer games and gamification. 
Developers need to be aware of how to best design such solutions to achieve high levels of engagement and enjoyment \cite{fensel2017contributing,kumar2013gamification}. 

\section{Study Design}
\label{studydesign}


We formulated the following research questions to guide our study:

\textit{RQ1: What are the range and nature of end user human-centric aspects that have to be addressed by software developers?} 


\textit{RQ2: How are different human-centric aspects addressed at different phases of software development?} 

\textit{RQ3: What current support is available to  developers and what improvement is needed?} 

\subsection{Survey and Interviews}

We designed an online survey targeted at a broad range of software developers and software team managers, to provide us with a big picture view of current practices, challenges and approaches being used to address key end user human-centric aspects in modern software development. The survey was composed of three sections: Demographic questions; Participant views on end user human-centric aspects ; and Particular techniques -- guidelines, practices, tools used to address diverse end user human aspects.
To complement this online survey we developed an interview protocol allowing us to drill down to more detailed information in one-on-one interviews. We wanted to selectively interview eligible developers to find out more insight on this topic. 


\subsection{Recruitment and Data Collection}

We ran our developer survey mid-2020 to late-2020. 
We recruited participants from our personal network, by advertising on LinkedIn and Twitter and through snowballing. We were particularly interested in surveying those developing software applications where (some of) the end users of the software have particular ``challenges'' e.g. physical, mental, age (very young or ageing), language proficiency, low socio-economic status, low access to technology and/or technology skills, and so on. 
We wanted to enhance the broad picture obtained from our survey to capture more specific information about development challenges and discuss these in detail with selected interviewees.  We recruited for the interviews from our own professional software developer networks, but also asked survey respondents to volunteer to be interviewed. We then selected from these contacts and volunteers a representative range of interviewees (domain of work, role, experience, etc). 
Originally we planned to conduct face-to-face interviews. Due to COVID-19 restrictions, all the interviews were conducted via Zoom. This allowed us to interview several participants from other countries and time zones. 

\subsection{Data Analysis}


Analysis of the quantitative data is mainly descriptive and explores common and uncommon aspects and key associations. Qualitative analysis included content analysis and thematic analysis. 
We identified key themes via open coding, grouped common themes and responses. We used closed coding for further analysis and found key themes.

\begin{figure*}[h]
\centering
\begin{minipage}{.5\textwidth}
  \centering
  \includegraphics[width=7.75cm,height=4.8cm]{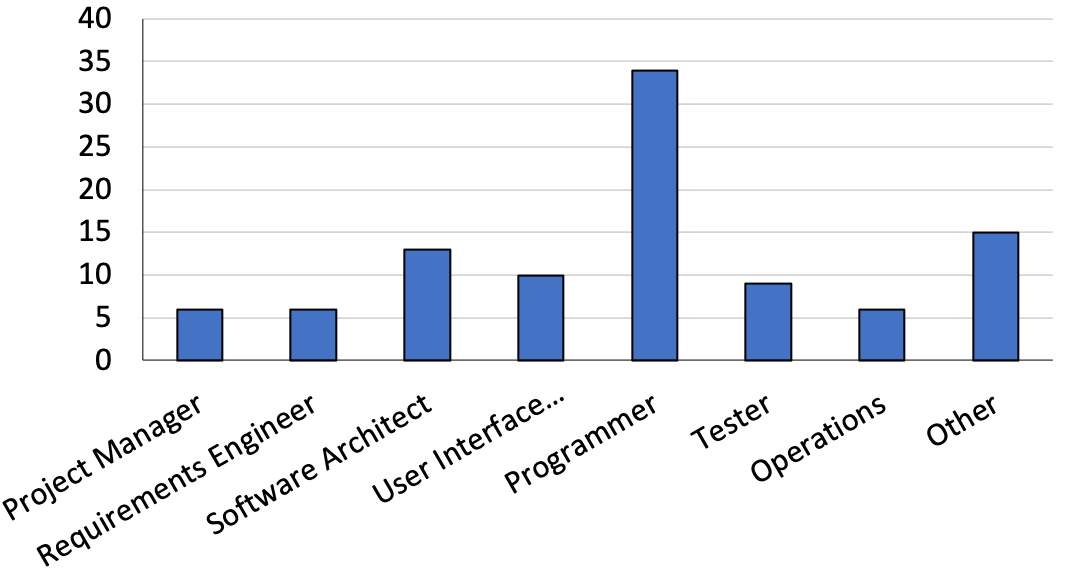}
  \captionof{figure}{Roles of survey respondents}
  \label{fig_roles}
\end{minipage}%
\begin{minipage}{.5\textwidth}
  \centering
  \includegraphics[width=7.75cm,height=4.8cm]{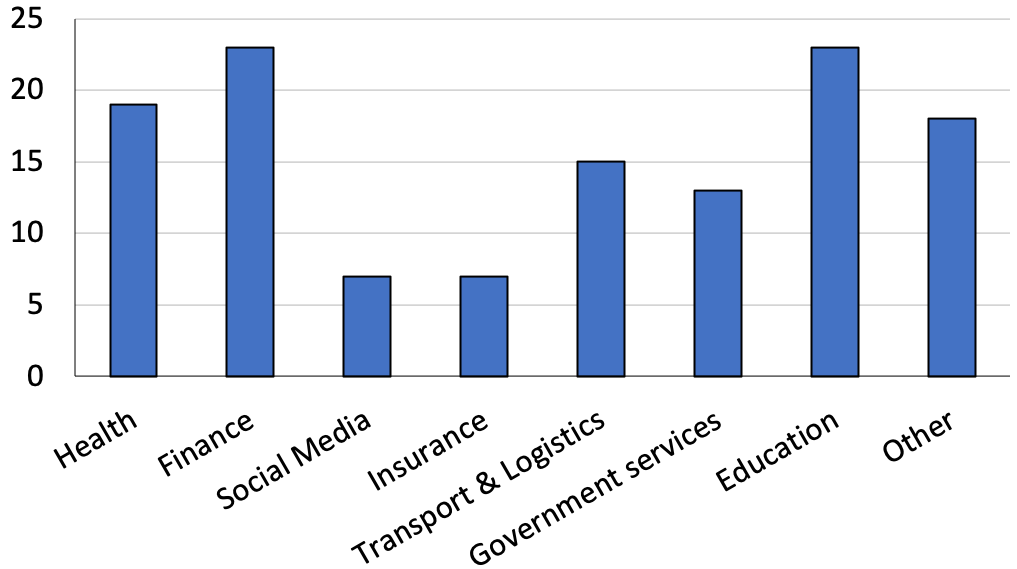}
 \captionof{figure}{Software domains worked in}
  \label{fig_domains}
\end{minipage}
\end{figure*}

\section{Results}
\label{sec:results}

\subsection{Participants}
We had over 130 online survey responses but only 60 were usable; the rest were removed from the final analysis due to incompleteness and/or poor quality of responses. Poor quality was decided where irrelevant or ``throw-away'' responses were encountered. Forty-four of these participants were male and 12 female; four did not state their gender. Ages ranged from 21-30 (23); 31-40 (18); 41-50 (8); 51-60 (6); 61+ (1), and under 20 (1); three did not state their age.
Years of experience ranged from 1-5 years (22); 6-10 years (12); 11-15 years (9); 16-20 years (6); 21-25 years (3); 26-30 years (2), 31-35 (1) and 36-40 (1); three not stated.
Most developers came from a Computer Science or Software Engineering training background - 22 (CompSci); 9 (SoftEng); 9 (IT/InfoSys); 5 (Computer Eng); others (one each) were from Robotics, Physics, Forensic Computing, AI/ML/Vision, and Neuroscience; the rest did not state their background. 
Figure \ref{fig_roles} shows current roles of our survey respondents, and Figure \ref{fig_domains} shows the different domains they have worked in. 

We also interviewed 12 respondents - 11 were male and one female (9 from Australia, 2 from New Zealand and 1 from the Middle East). Ages ranged from 21-30 (1); 31-40 (4); 41-50 (5); and 51-60 (2). 
Years of experience ranged from 1-5 years (1); 6-10 years (3); 11-15 years (1); 16-20 years (4); and 30+ years’ experience (3).
The interviewees covered a broad range of roles including project managers (4), requirements engineers (1), software architects (2), user interface designers (2), programmers (2), testers (1) and other (5); many people performed more than one role.

\begin{figure*}
    \centering
    \includegraphics[width=0.8\linewidth,height=3.75cm]{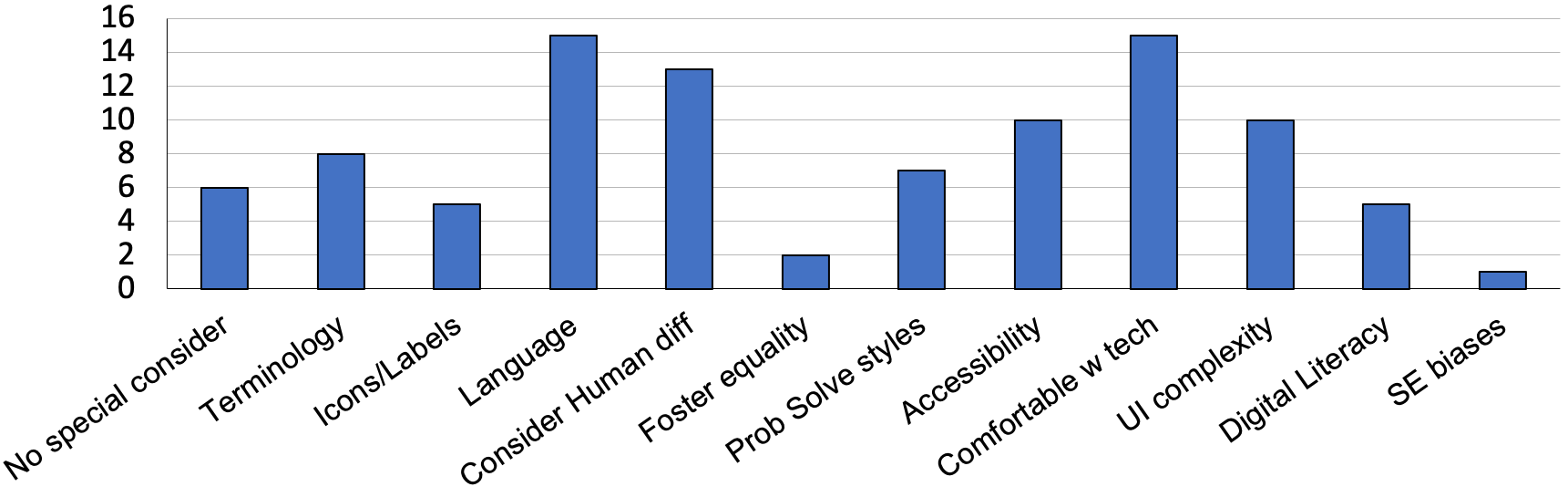}
    \caption{How these issues are taken into consideration in survey respondents work}
    \label{fig:reasons_critical.opng}
\end{figure*}

\subsection{Answers to Our Research Questions} 


\textbf{RQ1: What   are   the   range   and   nature   of   end   user human-centric aspects that have to be addressed by software developers?}

\begin{figure*}
    \centering
    \includegraphics[width=\linewidth,height=6.5cm]{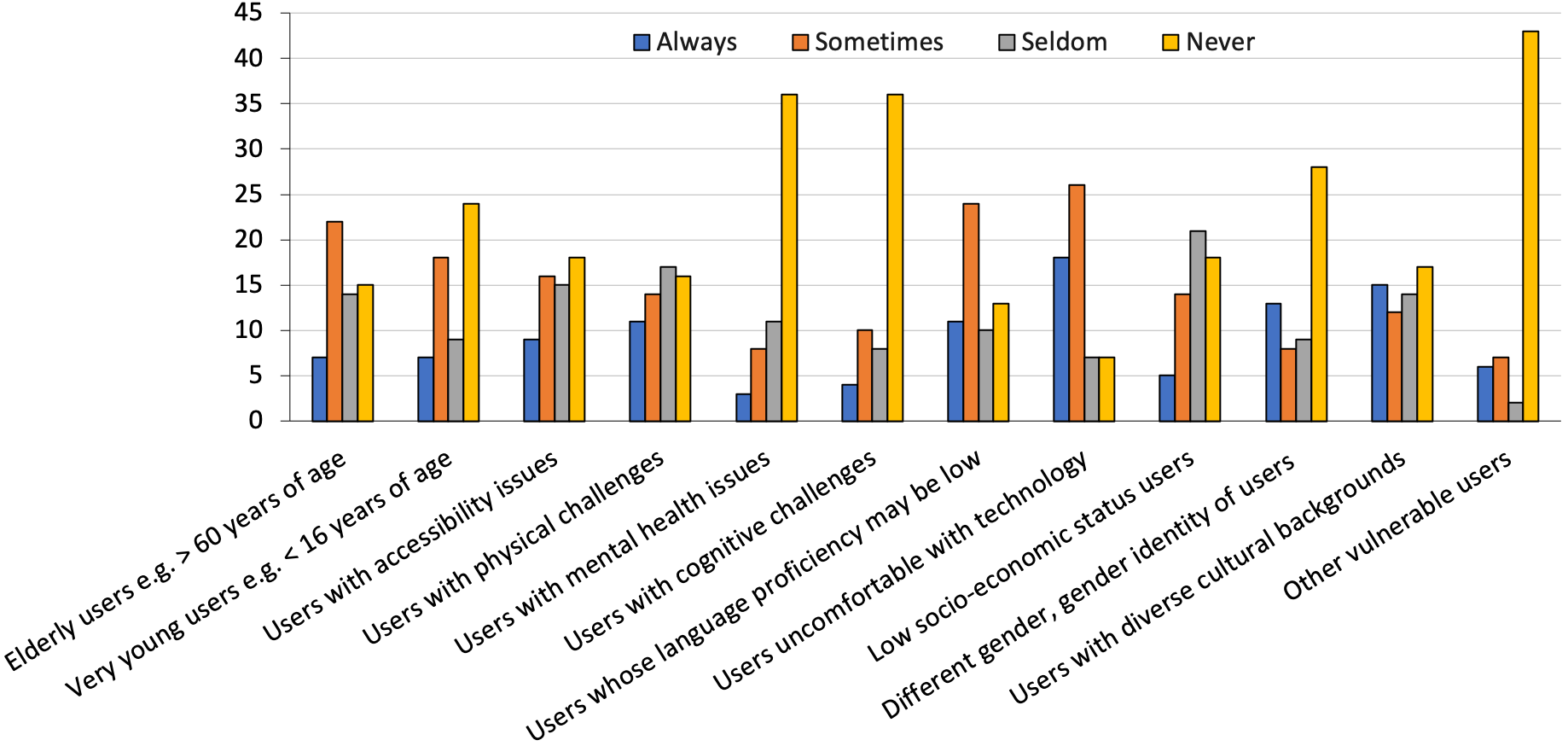}
    \caption{End user human-centric aspects survey respondents need to address}
    \label{fig:aspects_addressed}
\end{figure*}

We asked survey participants to tell us what end user human-centric aspects they have had to address in their software projects, summarised in Figure \ref{fig:aspects_addressed}. Ageing users, users with accessibility needs, those with physical challenges, those with language proficiency issues and uncomfortable with technology, and those with diverse cultural background were areas more highly reported. 
Interviewees described specific human-centric aspects they had to address and how they managed these challenges. Most common issues included ageing users, users who were technologically challenged, those who were from diverse cultural backgrounds and/or spoke languages other than English, specialised groups with unique work contexts, and even personality types.

\begin{figure*}
    \centering
    \includegraphics[width=0.9\linewidth,height=5.7cm]{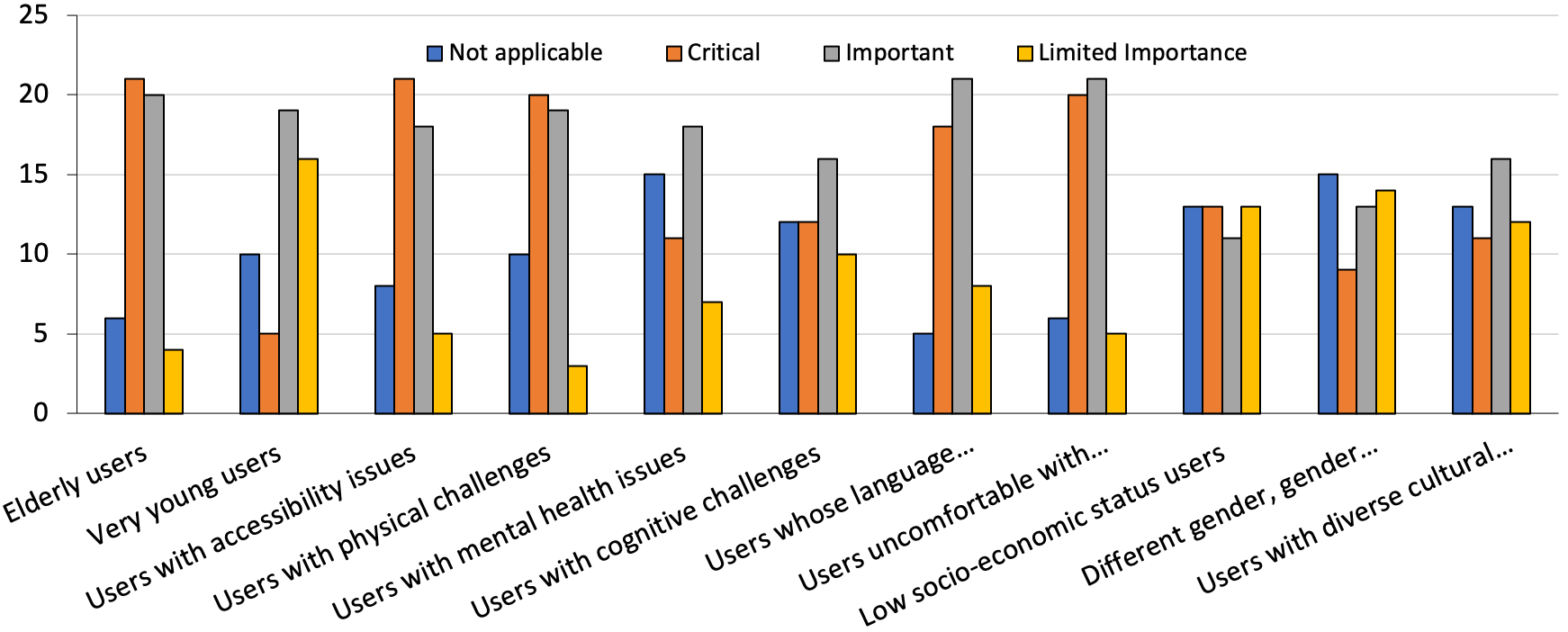}
    \caption{End user human-centric aspects survey respondents judge to be critical (or not) in their work}
    \label{fig:critical_aspects.png}
\end{figure*}

\textbf{Technical proficiency: }For example, developers had to adapt software for users with low technological capability: \textit{``there's a lot of [users] that struggle with digital technology, even to the point where we're actually building a web application that was previously just a mobile application just so that it's accessible to everyone."}

\textbf{Age: }Some addressed the issue of developing for users of different age groups:
\textit{``Many ticket officers/operators are middle-aged or even more senior. They're usually busy, less willing to explore the functionality of our software, and have to multitask."}, and 
\textit{``In our health systems we have a large group of users [who] are ``elderly". They include both clinical service providers (e.g. Doctors, nurses and service staffs etc.) and elderly patients"}.

\textbf{Culture: }Cultural differences were observed between the developers and the users:
 \textit{``You put the robot into the wild, you discover things you didn't foresee. We were very conservative, to try to not to offend anyone. But you still discover things, like, I did not expect this question: ‘Do you believe in God?’. People were very insistant on getting answers on this topic in this area. I think it is very important to have people who can think in this context in an early phase."} 


Some developers suggested that some domains tend to come with more end user human-centric aspects including health, financial and community apps, social media apps and safety-critical systems. 

\textbf{Clinical software: }In a clinical setting, to make sure the software was used according to the clinicians' needs, they used different panels for different parts of input for the clinician to click the panel that was being discussed, and if something else was suddenly being discussed, they could just switch to a different panel without skipping forward and backward to a few screens to get to the right spot. However, they
\textit{``did not really realise this up until they tried it in the clinics and the first version was trialled.”} The developers realised that \textit{``It didn't match up to the way people having conversations with the [doctor].”} Another issue they did not realise until it was tested was that they should not present the details related to the user’s cancer prediction in a same way to all the patients. 
\textit{``Because if a person has a risk calculator and their risk comes out to be in a high-risk category. If you just program the tool to present that risk to the person like anyone else. That \textbf{can really be stressful for the patient and it can induce anxiety}, which is, all the things we wanted to avoid in the goals of this project”}. 

\textbf{Games: } It is essential to address user  emotions, engagement, age and language. 

\textbf{Others: }In financial, community and social media application domains there is very wide range of end users with different expectations and needs.
Some developers flagged the issue of end users with multiple, interacting human-centric aspects that are very challenging to address:
    \textit{``It ties in with physical challenges (e.g. screen readability and its impact on deteriorating eyesight) and being comfortable using technology, which many older people are not"}. 
Some developers also noted they had little control over how many end user human-centric aspects were addressed, or even if they were addressed. Several organisations had dedicated UX teams, and larger ones ``customer experience" teams -- a development manager noted how their team didn't have direct access to end users and communication of end user needs or difficulties came through mixed channels.

\textbf{RQ2: How are different human-centric aspects are addressed at different phases of software development?}

We asked developers about the relative difficulty of addressing these end user human-centric aspects 
during one or more phases 
on a scale from 0 (no challenge) to 100 (most challenging). The survey and interviews reflected similar findings however, interview participants emphasized that human-centric aspects need to be better considered at all stages of development.

\textbf{Requirements engineering: }
Some of the key challenges stated included: gathering requirements for these end users is very hard (reported for elderly, young, mentally and physically challenged end users); addressing the issues this end user group has is critical for the software to be useful (accessibility, physically challenged, cultural differences); the team lacks sufficient knowledge how to address these issues (reported for several of these human-centric aspects); there are ethical issues in gathering these requirements (young children); finding and communicating with suitable end users with these challenges is difficult (reported for many of these human-centric aspects); a very wide range of issues for end users with this human-centric aspect (accessibility needs and ageing users); its hard to satisfy all end users with these human-centric aspects (accessibility, culture, language); the issue is very complex (cultural differences); and its hard to meet these requirements with suitable designs (reported for several  human-centric aspects). 
While some developers believed the requirements phase should be the most important, many acknowledged that it was not unusual for users to either expect too much and developers had to manage their expectations, or for users to change their mind, making it important to have checks in place throughout the development cycle.
One developer put it succinctly saying that even once a prototype was developed, it was still important to evaluate human-centric aspects as \textit{``Sometimes when we are starting to develop, they are only concepts. \textbf{We don't even know what kind of implications the technology may have}."}

\textbf{Design and development: }Key difficulty reasons reported for design and implementation-related tasks included: it was hard to find a balance between designs that met different needs (reported for elderly, young, accessibility); it takes a lot more effort to design solutions (accessibility); it is hard to include these end users in the design process (children); there are limited design tools exist to help (young and accessibility aspects); existing standards are hard to apply (accessibility); it is hard to know characteristics of users and their preferences (accessibility and gender); and it is hard to foresee the possible range of end user human issues (culture, accessibility, and language aspects).


\textbf{Testing and maintenance: }Key reasons given for difficulty for test and maintenance-related tasks included finding representative testers (reported for ageing users, those with cognitive challenges, different genders and cultural diversity); need for extensive testing (accessibility); difficulty of testing and fixing if the developers do not have these challenges themselves (many); difficulty determining who the end users are (for language and socio-economic status); difficulty determining specific actual usage issues (for young, accessibility, culture, language, socio-economic aspects); and potential ethical issues (for recruiting testers with mental health challenges).
Testing was seen as a problem if testers were not using the software in the context it was designed for e.g. under stress:
\textit{``when they're just doing stuff, they have a very different behaviour than when they're stressed, you have to be much more clear, it's really important for that testing under that real life sort of situation, testing under the worst case,  because it's really important that the behaviour under all those work conditions is taken into account."}




\textbf{RQ3: What support is available to the developers and what improvement is needed?}

\begin{figure*}[h]
\centering
\captionsetup{justification=centering}
\begin{minipage}{.6\textwidth}
  \centering
  \includegraphics[width=11cm,height=7.5cm]{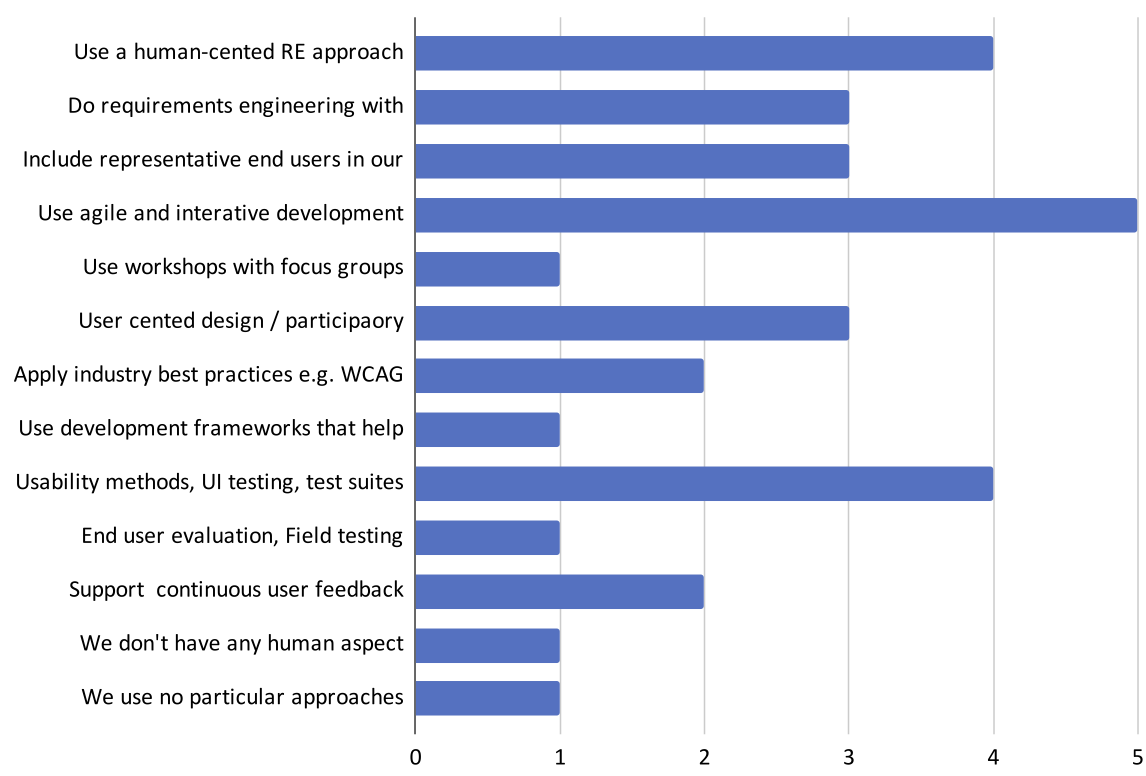}
  \captionof{figure}{Techniques used to address end user human-centric aspects} 
  \label{fig_techniques_used}
\end{minipage}%
\begin{minipage}{.38\textwidth}
  \centering
  \includegraphics[width=6cm,height=7.5cm]{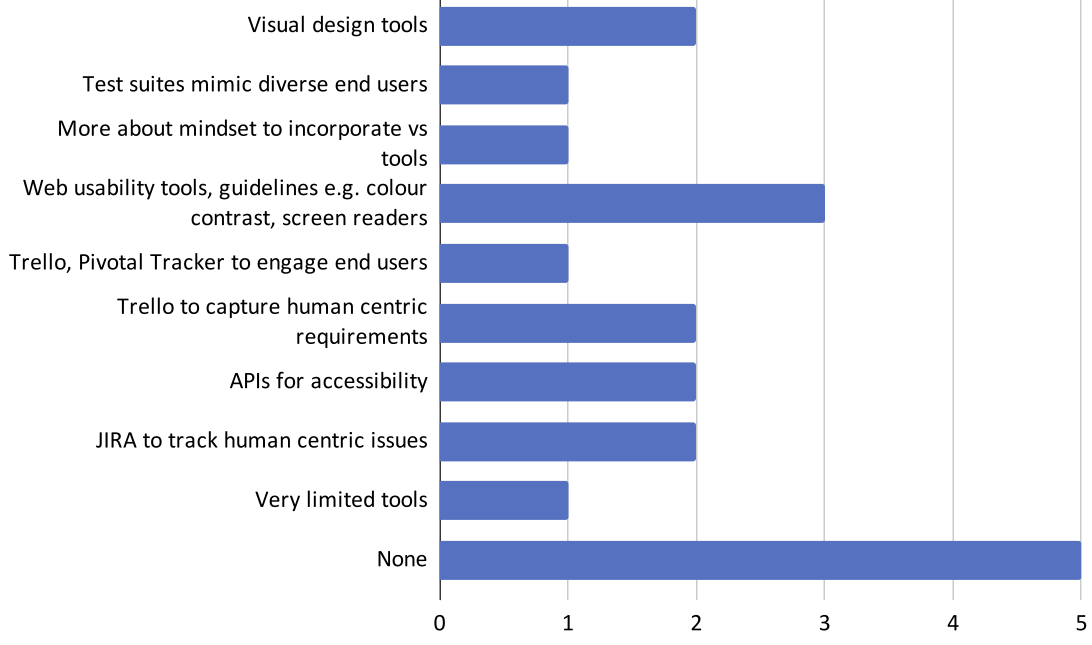}
 \captionof{figure}{Tools to address  human-centric aspects}
  \label{fig_tools_used}
\end{minipage}
\end{figure*}


We asked survey respondents to tell us what key techniques and tools, if any, they currently or have used to address some of these end user human-centric aspects in their project work. Figure \ref{fig_techniques_used} summarises key techniques used. Surprisingly few reported using ``human-centred" RE and Design approaches. Agile and iterative software development methods and usability evaluation techniques were claimed to be beneficial by several. A few reported feedback mechanisms, ``best practices" such as applying standards, and including end users in the process were all critical.
A few use standards/guidelines for specific human-centric aspects, especially usability and those with physical or cognitive challenges. A few use visual design tools to model user human-centric aspects, JIRA to track human-centric aspect-related defect fixing; Trello cards to capture human-centric requirements, and accessibility APIs. Interviewed developers reported using tools including Jira, Zendesk, DevOps, Nagios and Selenium to increase communication between developers, stakeholders and users. Similarly to the survey results, many talked about successfully using agile and iterative software development methods, participatory scrum, brainstorming techniques, and methods for increasing rapid feedback to increase capturing and addressing human-centric aspects. One team even trained users to be scrum masters. Smaller teams and solo practitioners employed informal methods for getting feedback and resolving problems e.g. email and spreadsheets.

We asked developers what improved tools and techniques they thought would help them. Key examples given included better development processes to improve target end user collaboration; better guidelines and practices to follow to address diverse end user human-centric aspects in software; better requirements capture and human-centric aspect modelling support that would enable them to identify and better track these end user needs throughout development; AI-based tools to automatically advise on missing end user human-centric aspects e.g. to prompt them to consider certain end user human-centric aspects in different situations; and more ``live" or \textit{in situ} testing with representative end users, to get richer feedback on issues that arise in software from lack of consideration of end user human-centric aspects. A number of other suggestions were given including a need for better education of software engineers about diverse end user human-centric aspects and their impact on software usage; simpler GUIs for many end user populations; better defect reporting to enable diverse end users to more easily identify, describe and report problems they have with their software and so on. 


Participants also suggested developers trying out  being ``users":
 \textit{``I think if you take two days out of a development cycle and send half of your developers to be the user for a couple of days, you'll pay that. You'll save that in tons later on that project."}


\subsection{Limitations}
\label{sec:ThreatsValidity}

Ideally we would have had a larger number of survey respondents and interviewees. The demographics of the respondents did however give us a reasonable spread of experience, domain and gender. We purposively chose interviewees from those who volunteered to give us a broad range of demographics.
Our survey 
questions may have been misinterpreted by some respondents and some may have not taken due care with the 
survey. 
We did our best to use terminology and brief explanations in the survey that developers would correctly interpret based on a pilot run. 

\section{Summary}
\label{sec:conclusion}

We reported results of online survey and interviews of software engineers exploring challenges they face in addressing a range of human-centric aspects of their end users. Most software engineers share few human-centric aspect characteristics with some of their end user groups. Key challenges identified included a lack of education, knowledge, experience, guidelines and tools about ways to best address some end user human-centric aspects; difficulty in recruiting representative end users and working with them throughout development; sheer difficulty in addressing a wide range of sometimes conflicting human-centric aspects of end users; inability to satisfy all potential end users with differing human-centric aspects; and lack of time, budget and management support in addressing many end user human-centric aspects. We want to carry out observational studies  with a small number of software teams to observe developers working on software to see how they discuss and address these issues. We also want to survey and selectively interview a range of stakeholders and end users of software applications to better understand their challenges using the software. We want these learnings to help us to trial with developers and end users new software engineering processes, techniques and tools to address (some of) the challenging, outstanding issues in human-centric aspects in software for end users. 
\section*{Acknowledgements}
All the authors are supported by ARC Laureate Fellowship  FL190100035. 

\bibliographystyle{plain}
\bibliography{main}
\end{document}